\documentclass[column,aps,amsmath,showpacs,floatfix,superscriptaddress,nofootinbib,plb]{revtex4}
\usepackage{graphicx}% Include figure files
\usepackage{dcolumn}% Align table columns on decimal point
\usepackage{bm}% bold math
\usepackage{ulem}% bold math
\usepackage{overpic}
\usepackage{amssymb}
\usepackage{amsmath}
\usepackage{datetime}

\newcommand{\nn}{\nonumber}

\def\beq{\begin{equation}}
\def\eeq{\end{equation}}

\usepackage{soul}
\usepackage{cancel}

\usepackage[usenames]{color}

\begin{document}

% \ln \rho\int_{0}^{\infty}\frac{dk k J_{0}(kx_{10})}{k}]\\
 \title{%Small-$x$ asymptotics of the unintegrated dipole gluon distribution %function
Unintegrated dipole gluon distribution at small transverse momentum}
% 
% 
% 
%  \author{Raktim Abir and Mariyah Siddiqah}   
% %
% %\affiliation{Department of Physics, Aligarh Muslim University, Aligarh - $202002$, UP, India.}  
% 
\author{Mariyah Siddiqah, Nahid Vasim, Khatiza Banu and Raktim Abir \\  Department of Physics, Aligarh Muslim University, Aligarh - $202002$, India. 
\\ [2.0ex] Trambak Bhattacharyya \\ UCT-CERN Research Centre and Department of Physics, R W James Building, University of Cape Town, Rondebosch 7701, Cape Town, South Africa.}

%\author{Nahid Vashim} 

%\author{Khatiza Banu} 
%
%\author{Raktim Abir}   
%
%\affiliation{Department of Physics, Aligarh Muslim University, Aligarh - $202002$, UP, India.}  

%\author{Trambak Bhattacharyya \\ UCT-CERN Research Centre and Department of Physics, R W 
%                  James Building, University of Cape Town, Rondebosch 7701, 
%                  CapeTown, South Africa.} 

%\affiliation{UCT-CERN Research Centre and Department of Physics, R W 
%                  James Building, University of Cape Town, Rondebosch 7701, 
%                  CapeTown, South Africa.}

%\affiliation{Department of Physics, Aligarh Muslim University, Aligarh - $202002$, UP, India.}  

% 
%\author{Raktim \surname{Abir}} 
% \email{raktimabir@gmail.com}

%\author{Trambak Bhattacharyya} 
% \email{trambak.bhattacharyya@gmail.com} 
 
%\author{Mariyah Siddiqah} 
% \email{shah.siddiqah@gmail.com}
 
%\author{Nahid Vashim} 
% \email{vasim.nahid19@gmail.com} 

%\author{Mariyah Siddiqah$^{1}$,  Nahid Vashim$^{1}$, Raktim Abir$^{1}$ and Trambak Bhattacharyya$^{2}$}  

 %\affiliation{$^1$Department of Physics, Aligarh Muslim University, Aligarh - $202002$, UP, India,}

 %\affiliation{$^2$UCT-CERN Research Centre and Department of Physics, R W 
 %                 James Building, University of Cape Town, Rondebosch 7701, 
 %                 CapeTown, South Africa.}
 
%\affiliation{$^1$Theory Division, Saha Institute of Nuclear Physics  1/AF Bidhannagar, Kolkata 700064, India.}

% 
 
% 

% 
%  
  \begin{abstract}
%We derive analytical results for the small $x$ asymptotic behavior of the unintegrated dipole gluon distribution function. We studied the Levin-Tuchin solution of the Balitsky-Kovchegov equation, valid in the asymptotic black disc limit, to derive the results in the form of a series of Bell polynomials. 
We derive analytical results for  unintegrated color dipole gluon distribution function at small transverse momentum.
% This study have been done only at asymptotically high energy limit or black-disk limit. 
By Fourier transforming the $S$-matrix for large dipoles
%as given by the Levin-Tuchin solution of the Balitsky-Kovchegov equation, to transverse momentum space, 
we derive the results in the form of a  series of Bells polynomials. Interestingly, when resumming  the series in leading log accuracy, the results showing up striking similarity with the Sudakov form factor with role play of coupling is being done by a constant that stems from the saddle point condition along the saturation line.   
  \end{abstract}
 \date{\today ~~\currenttime}

 \maketitle
 
 \section{introduction}
The discovery of rapidly growing cascade of gluons and sharp non-linear rise of its distributions in DIS experiments at HERA collider provides indirect experimental evidence that proton at high energy is a hugely complex many body quantum system where gluons are the dominant  degrees of 
freedom. 
% C:
%
%
By now many studies have been performed in the theoretical front to develop frameworks that extend our understanding on the structure of proton beyond just one dimensional ordinary parton distribution functions (PDFs).
The attempts are mostly based on considering other (than regular PDFs) relatively closer descendants of the original, yet unknown, Wigner distribution functions which presumably contain all the informations.  
%C:
%
%
Transverse momentum dependent parton distributions (TMDs) or unintegrated parton distribution functions (UPDFs) are such examples that provide, in addition to longitudinal momentum fraction $x$ of the parton, details of transverse momentum distribution and therefore contain much more detailed information on the internal structure of protons relative to the ordinary PDFs \cite{Bacchetta:2016ccz}.   \\
%C:
%
%

On the experimental front the unintegrated gluon distribution functions are among the key topics to be fully investigated at current and future electron-ion collider facilities including JLab's 12 GeV Upgrade, eRHIC and the planned EIC.
The focus would be on both polarised or unpolarised parton distributions
for both  spin polarized and unpolarised  protons. 
% C:
%
%
It has been anticipated that the unintegrated gluon distribution  can either be directly probed in the quark-antiquark jet correlation in deep inelastic scattering (for unintegrated Weizs\"{a}cker-Williams gluon distributions) or  in the direct photon-jet correlation in $pA$ collisions (for unintegrated dipole gluon distributions).  
% c:
%
%
The unintegrated gluon and quark distributions involved in other different processes, including other dijet channels in $pA$ collisions, are actually related to this two widely proposed ones: the Weizs\"{a}cker-Williams gluon distribution and the dipole gluon distribution in the large-$N_c$ limit \cite{Dominguez:2011wm,Buffing:2013eka,Hatta:2016dxp,Xiao:2017yya}.   \\
% C:
%
%

The deep inelastic scattering experiments at HERA also provide intense indications that there exists a novel, yet unexplored, saturation regime  in high energy limit of QCD where the many-body dynamics inside the proton is intrinsically non-linear in character.  In this regime the cascading gluons occupy the phase space in the final state to such an extent that fusion of newly formed gluons begin to start leading to the origin of gluon saturation with a characteristic momentum scale $Q_s$\cite{Gribov:1984tu}. 
This fusion of multiple gluons to single gluon eventually restore the  unitarity of the scattering $S$-matrix, which will otherwise violated by the almost exponential growth of gluon multiplicity. This was first studied by Balitsky \cite{Balitsky:1995ub,Balitsky:2001gj} within Wilson line formalism  leading to a hierarchic chain formed by the Wilson line operators and later by Kovchegov \cite{Kovchegov:1999yj,Kovchegov:1999ua} in the Mueller's color dipole approach \cite{Mueller:1993rr,Mueller:1994jq,Chen:1995pa} where the Balistky  hierarchic chain of Wilson line operators reduced to a closed form equation in the large-$N_c$ limit.  The Balitsky-Kovchegov (BK) equation  is a integro-differential equation,  the integral kernel for both linear and non linear terms are happen to be identical, and has a simple interpretation of one parent color dipole in the initial state splitting into two daughter dipoles in the final state. 
Since then a lot of progress has been made in solving the BK equation both analytically and numerically and until now only a few limiting analytical solutions of leading order BK equation exits.   \\

In this article we consider Levin-Tuchin (LT) solution \cite{Levin:1999mw,Levin:2000mv} of the LO BK equation and derive unintegrated dipole gluon distribution, for small transverse momentum, in the form of a series of Bells polynomials. Within the leading log approximation the series resummed as,  
  \begin{eqnarray}
 \left. x G^{DP}\left(x,k_{\bot}\right)\right|_{Q_s \gtrsim k_\perp  \gg \Lambda_{\rm QCD}}^{} 
 &\approx &-\frac{S_{\bot}N_c \tau}{\pi^3\alpha_{s}}~\ln\left(\frac{k_\perp^2}{4Q_s^2(Y)}\right) \exp \left[-\tau \ln^2\left(\frac{k_\perp^2}{4Q_s^2(Y)}\right) \right]~, \label{main1234}
 \end{eqnarray} 
%
%  \begin{eqnarray}
 %x G^{DP}(x,k_{\bot})
% &=&\frac{S_{\bot}N_c}{2\pi^3\alpha_{s}}~\cal{F}'\left(\beta\right) \exp \left[\cal{F}(\beta)\right]~,
 %\end{eqnarray} 
 %where, 
 %\begin{eqnarray}
%{\cal{F}}(\beta) &=&   -\tau \beta^2-2 \sum_{j=0}^{\infty}\frac{\zeta(2j+1)}{2j+1}~\tau^{j+1/2} ~H_{2j+1}\left(\sqrt{\tau}\beta\right)~, 
 % \end{eqnarray}  
 % and, 
 % \begin{eqnarray} 
%{\cal{F}'}(\beta) \equiv \frac{\partial {\cal F} }{\partial \beta}  &=&  - 2\beta\tau  - 4 \sum_{j=1}^{\infty}\zeta(2j+1)~\tau^{j+1}
% ~H_{2i}\left(\sqrt{\tau}\beta\right) ~,
% \end{eqnarray}  
%with, 
% \begin{eqnarray}
 %\beta \equiv \ln (k_\perp^2/4Q_s^2). 
% \end{eqnarray}
where $k_\perp$ is transverse momentum of the parton, rapidity variable $Y\equiv \ln(1/x)$ and 
%\begin{eqnarray}
%\tau= \frac{1+2i\nu_0}{4\chi\left(0,\nu_0\right)} \approx 0.2 ~,  
%\end{eqnarray}
$\tau  \approx 0.2$ a constant that stems from the saddle point condition along the saturation line.  As Levin-Tuchin (LT) solution is valid only for larger dipoles $ x_\perp  \gtrsim 1/Q_s $ deep inside the saturation region Eq.(\ref{main1234}) is expected to be valid in the relatively smaller momentum range $Q_s \gtrsim k_\perp  \gg \Lambda_{\rm QCD}$. Employing McLerran-Venugopalan (MV) \cite{McLerran:1993ni,McLerran:1993ka,McLerran:1994vd} model or phenomenological Golec-Biernat and M.~Wusthoff (GBW)  \cite{GolecBiernat:1999qd,GolecBiernat:1998js} form of $S$-matrix, we also derive the UDGD function as, 
\begin{eqnarray}
\left. x G^{DP}\left(x,k_{\bot}\right)\right|_{ k_\perp \gtrsim Q_s  \gg \Lambda_{\rm QCD}}^{Y \sim 0} 
 &\approx &\frac{S_{\bot}N_c}{2\pi^3 \alpha_{s}}~\frac{k_\perp^2}{Q_s^2(Y)}~\exp\left(-\frac{k_\perp^2}{Q_s^2(Y)}\right), 
\label{xG-DP1001111} 
 \end{eqnarray}
 which is expected to be valid around $Y \sim 0$. We further discussed the distribution in the extended geometric scaling region (just outside the saturation boundary) and also for very large momentum where power-law fall with increasing transverse momentum.     \\

The article is organised as follows: In Sec.II we review the operator definition of unintegrated gluon distribution functions, the WW distribution and color dipole distribution, and their possible connection to $S$-matrix at small $x$. We then adopt  form of $S$-matrix valid at different kinematic ranges outside or in the vicinity of saturation boundary.  In Sec. III we derived our main result for the unintegrated color dipole gluon distribution at small momentum. Finally we conclude in Sec.III.

\section{Unintegrated gluon distributions and small x dynamics}
Both types of universal unintegrated gluon distribution functions: Weizs\"{a}cker-Williams (WW) distribution and dipole gluon distribution are essentially dimension four two point correlation function of classical gluon fields (non-abelian Weizs\"{a}cker-Williams fields) of relativistic hadrons. The operator definition of Weizs\"{a}cker-Williams gluon distribution is, 
 \begin{eqnarray}
 x G^{WW}\left(x,k_{\bot}\right)=2\int\frac{d\xi^{-}d^2\xi_{\bot}}{(2\pi)^3P^{+}}e^{ixP^{+}\xi^{-}-ik_{\bot}.\xi_{\bot}}
 ~\langle P|{\rm Tr}\left[F^{+i}(\xi^{-},\xi_{\bot})~\mathcal{U}^{[+]\dagger}~F^{+i}(0,0)~\mathcal{U}^{[+]}\right]|P\rangle~, 
 \label{WW-Operator}
 \end{eqnarray}
 whereas the operator definition of color dipole gluon distribution in the fundamental representation is, 
 \begin{eqnarray}
 x G^{DP}\left(x,k_{\bot}\right)=2\int\frac{d\xi^{-}d^2\xi_{\bot}}{(2\pi)^3P^{+}}e^{ixP^{+}\xi^{-}-ik_{\bot}.\xi_{\bot}}
 ~\langle P|{\rm Tr}\left[F^{+i}(\xi^{-},\xi_{\bot})~\mathcal{U}^{[-]\dagger}~F^{+i}(0,0)~\mathcal{U}^{[+]}\right]|P\rangle ~.
 \label{DP-Operator}
\end{eqnarray}
In both the definitions $F^{\mu\nu}$ is gluon field strength tensor $F^{\mu\nu}_a$ 
and the gauge links involved are,   
%
%
%  \begin{eqnarray}
%  \mathcal{U}^{[+]}&=&U^n\left[0,\infty;0_{\bot}\right]U^t\left[\infty;0_{\bot},   \infty_\perp\right]
%                    U^t\left[\infty;\infty_{\bot},\xi_\perp\right]U^n\left[\infty;\xi^-,\xi_\perp\right] \\
%  \mathcal{U}^{[-]}&=&U^n\left[0,-\infty;0_{\bot}\right]U^t\left[-\infty;0_{\bot},\infty_\perp\right]
%                    U^t\left[\infty;\infty_{\bot},\xi_\perp\right]U^n\left[-\infty;\xi^-,\xi_\perp\right]
%  \end{eqnarray}
%
\begin{eqnarray}
 \cal{U}^{[+]} &=&
 U^n\left[0^{-},0_{\bot};\infty^{-},0_{\bot}\right]U^t\left[\infty^{-},0_{\bot};\infty^{-},\infty_\perp\right]
                   U^t\left[\infty^{-},\infty_{\bot};\infty^{-},\xi_\perp\right]U^n\left[\infty^{-},\xi_{\bot};\xi^-,\xi_\perp\right]~, \\
\cal{U}^{[-]}&=&U^n\left[0^{-},0_{\bot};-\infty^{-},0_{\bot}\right]U^t\left[-\infty^{-},0_{\bot};-\infty^{-},\infty_\perp\right]
                   U^t\left[-\infty^{-},\infty_{\bot};-\infty^{-},\xi_\perp\right]U^n\left[-\infty^{-},\xi_{\bot};\xi^-,\xi_\perp\right]~, 
 \end{eqnarray}
where the longitudinal ($U^{n}$) and transverse ($U^{t}$) gauge links are defined as, 
 \begin{eqnarray}
  U^n\left[a^{-},x_{\bot};b^{-},x_{\bot}\right]&=&\mathcal{P} \exp\left[ig~\int_{a^{-}}^{b^{-}}dx^{-} 
   A^{+}\left(0,x^{-},x_{\bot}\right)\right]      ~, \nn\\   
  U^t\left[x^{-},a_{\bot};x^{-},b_{\bot}\right]&=&\mathcal{P} \exp\left[ig~\int_{a_{\bot}}^{b_{\bot}}dx_{\bot} .~
   A_{\bot}\left(0,x^{-},x_{\bot}\right)\right] ~.\nn 
 \end{eqnarray}
 In some light-cone gauge insertion of transverse gauge links are mandatory in order to maintain the gauge invariance of the distribution functions. \\
 
 Now one may assume for simplicity that the nucleus is a cylinder with its axis along the $z$-axis, so that saturation momentum $Q_s$ does not depend on the impact parameter ($b_\perp$) (depends however on $x$) and integration over $b_\perp$ can be carried out simply by multiplying
 the integrand by the transverse area ($S_\perp$). It can then be shown that
 under this assumption, the color dipole gluon distribution can be written as \cite{JalilianMarian:1996xn}, 
 %
 %\begin{eqnarray}
 %x G^{WW}(x,k_{\bot})=\frac {S_{\bot}}{\pi^2\alpha_{s}}\frac{N_{c}^{2}-1}{N_{c}}\int \frac{d^2 r_{\bot}}{(2\pi)^2}
% \frac{e^{-i k_{\bot}.r_{\bot}}}{r_{\bot}^{2}}N\left(x, r_\perp\right)~~.
 %\end{eqnarray}
 %
 \begin{eqnarray}
 x G^{DP}(x,k_{\bot})&=&\frac{S_{\bot}N_c}{2\pi^2\alpha_{s}}~k_\perp^2
 \int\frac{d^2r_{\bot}}{(2\pi)^2}e^{-ik_{\bot}.r_{\bot}}
 ~\frac{1}{N_c}\langle{\rm Tr}~ U(r_\perp){\it U}^{\dagger}(0)\rangle_{x} ~ \nn \\
 &=&\frac{S_{\bot}N_c}{2\pi^2\alpha_{s}}~k_\perp^2
 \int\frac{d^2r_{\bot}}{(2\pi)^2}e^{-ik_{\bot}.r_{\bot}} S(x, r_\perp)~, 
 \label{xG-DP1001111111111} 
 \end{eqnarray}
where $S(x, r_\perp)$ is the $S$-matrix for the quark (anti-quark) dipole, with transverse separation of poles being $r_\perp$, scattering on a nuclear target with some high energy corresponding to Bjorken scaling variable $x$.  

Assumption of color dipoles as the degrees of freedom possibly be the most convenient way to study high energy evolution in QCD. 
Originally proposed by Mueller, and developed in transverse coordinate space,
it is easier to include the saturation effects in the model. 
%
%Typically one starts with a quark (anti-quark) pair in order to calculate probability of emission of a soft gluon off this pair. 
%Both the quark and anti-quark are to follow light cone trajectories and emitted gluons are calculated in the eikonal approximations.
 %
 %Adding contributions  coming from the quark and anti-quark together with their interference one gluon part of onium wave function found to be proportional to following integral kernel convoluted over the onium wave function 
 %with no soft gluon \cite{Mueller:1993rr},   
 %
 %
In this framework of  Mueller dipole model the non-linear  evolution of the $S$-matrix, ${S}(x, r_\perp :=x_\perp-y_\perp)$, is governed  
 by the Balitsky-Kovchegov equation (BK), in large-$N_c$ limit, as,   
 %
% \begin{eqnarray}
% \frac{\partial {S}(y_\perp,y_\perp';Y)}{\partial Y} &=& - \frac{\alpha_s N_c}{2\pi^2} 
%               \int d^2z_\perp \frac{(x_\perp-y_\perp)^2}{(x_\perp-z_\perp)^2(z_\perp-y_%\perp)^2}           \left[{S}(x_\perp , y_\perp;Y)-{S}(y_\perp,z_\perp;Y){S}(z_\perp,y_\perp,Y) \right]~.
% \label{suchitra}
% \end{eqnarray}
 The Balitsky-Kovchegov (BK) equation  takes into account all long-lived soft gluon emissions, off almost onshell hard quark, with the lifetime of the gluon usually much longer than the size of the nuclear target.  In the following we will take 
 %(1) a specific form of S-matrix, that acts as the initial condition (MV initial condition) for the BK equation, expected to be valid for some initial rapidity both inside and  outside the saturation region,  and (2) one particular asymptotic solution (LT solution) of the BK equation, valid deep inside saturation region, 
different limiting solutions of the above equation in the form of the $S$-matrix to derive $x G^{DP}(x,k_{\bot})$  appropriate for corresponding momentum ranges.

%In the following form of the $S$-matrix  at different kinematic regions of interest  in Eq.(\ref{xG-DP1001111111111}) to derive $x G^{DP}(x,k_{\bot})$ appropriate for corresponding momentum ranges. In the next two sub sections we will review some of the results that have been already reported elsewhere. 

\subsection{UDGD outside or in the vicinity of the saturation boundary}
The region with momentum $k_\perp \lesssim Q_s(Y)$ (corresponding to $r_\perp \gtrsim 1/Q_s(Y)$), where the non-linear effects become important, is the saturation region.
When the transverse momentum is high enough, $(k_{\perp} \gg Q_{s})$, it is well known from pQCD that {\cite{Kharzeev:2003wz,Marquet:2016cgx}), 
\begin{eqnarray}
\left. x G^{DP}(x,k_{\bot})\right|_{ k_\perp \gg Q_s} \propto \frac{1}{k_\perp^2}
 \label{xG-DP1001113} ~. 
\end{eqnarray} 
This is identical for both the unintegrated Weizs\"{a}cker-Williams distribution and color dipole distribution. \\

 In the vicinity but just outside of the saturation 
 boundary, $k_\perp \gtrsim Q_s(Y)$ (corresponding to $r_\perp \lesssim 1/Q_s(Y)$), dipole amplitude takes the form, 
 \begin{eqnarray}
N(r_\perp,Y)=1-S(r_\perp,Y)\propto \left[r_\perp Q_s(Y)\right]^{1+2i\nu_0}\approx \left[r_\perp^2 Q_s^2(Y)\right]^{\gamma_{\rm cr}}~,
 \end{eqnarray}
where $\gamma_{\rm cr}\sim 0.63$~. When substituting back to Eq.(\ref{xG-DP1001111111111}) UDGD shows expected power law fall in $k_\perp$ as \cite{Marquet:2016cgx}, 
\begin{eqnarray}
\left. x G^{DP}\left(x,k_{\bot}\right)\right|_{ k_\perp \gtrsim Q_s}
 &\propto &\frac{S_{\bot}N_c}{2\pi^3 \alpha_{s}}~\left(\frac{Q_s^2(Y)}{k_\perp^2}\right)^{\gamma_{\rm cr}}~.
 \label{xG-DP1001112} 
 \end{eqnarray}

\subsection{UDGD in GGM, MV and GBW models}

%
%Even though the integral is clearly dominated by the range where LT solution could be valid one should be prudent when interpreting the result in the domain of relatively large high $k_\perp$.  One may not trust Eq. (\ref{main1234})  around $k_\perp \approx \Lambda_{\rm QCD}$ (correspond to large $r_\perp \sim \Lambda_{\rm QCD})$ where perturbation theory breaks and also not for $k_\perp>Q_s(Y)$, corresponds to small dipole $r_\perp \sim \Lambda_{\rm QCD})$ contribution from which have been ignored as mentioned before. 
%
%

Kinematic domains where in-medium interactions can be approximated by stochastic multiple soft gluon scatterings the scattering  matrix appears to be a Gaussian in dipole size. Outside the saturation region and for small dipole $r_\perp \ll 1/Q_s$  the scattering matrix in the Glauber-Gribov-Mueller (GGM) model found to scales with $r_\perp$ as,  
\begin{eqnarray}
S(r_\perp, Y\sim 0)= \exp\left(-\frac{1}{4}r_\perp^2 Q_{s0}^2\right)\approx 1- \frac{1}{4}r_\perp^2 Q_{s0}^2 ~,    \label{GGM}  
\end{eqnarray} 
 Since in a zero-size dipole the color charges of quark and the anti-quark cancel each other leading to the disappearances of interactions with the target  (the color transparency effect), at small $r_\perp$ we have $N=1-S\sim r_\perp^2$, so the amplitude $N$ is zero for zero dipole size. Even in McLerran-Venugopalan model or phenomenological Golec-Biernat and M.~Wusthoff model the $S$-matrix takes the following form, 
 \begin{eqnarray} 
 S_{\rm MV}(r_\perp, Y)=1-N(r_\perp,Y)=\exp\left(-\kappa r^2_{\perp}Q_s^2(Y)\right)~,
 \label{MV}
 \end{eqnarray}
 where $N$ is the imaginary part
 of the dipole-nucleus amplitude  and $\kappa\sim 1/4$  usually fixed from the definition of the saturation scale $Q_s$.  Eq.(\ref{MV})  is expected to be valid at some low initial rapidity both little inside and outside the saturation region and often used as initial condition for the full $Y$ evolution of $S$-matrix through the Balitsky-Kovchegov equation. 
The $S$-matrix in both Eq.(\ref{GGM}) and Eq.(\ref{MV}) is Gaussian in the  variable $x_\perp Q_s(Y)$ with a (model dependent) variance. Therefore substituting Eq.(\ref{MV}) in Eq.(\ref{xG-DP1001111111111}) and evaluating the integral is easy leading to a {\it Gaussian} $\bigotimes$ {\it quadratic} distribution of transverse momentum as \cite{Petreska:2015rbk,Marquet:2016cgx} (See also \cite{Abir:2015qva}), 
 \begin{eqnarray}
 x G^{DP}(x,k_{\bot})
 &=&\frac{S_{\bot}N_c}{2\pi^3 \alpha_{s}}~\frac{k_\perp^2}{Q_s^2(Y)}~\exp\left(-\frac{k_\perp^2}{Q_s^2(Y)}\right)~.
 \label{xG-DP100} 
 \end{eqnarray}

%\subsection{S-matrix for large dipole ($r_\perp \gtrsim 1/Q_{s}(Y))$ at large rapidity $Y \gg 1$}
%
 
\section{UDGD deep inside the saturation boundary} 
 Deep inside saturation region one may take Levin-Tuchin form of S-matrix valid for larger dipoles $ x_\perp  \gtrsim 1/Q_s $ therefore could be appropriate to explore relatively smaller momentum range $Q_s \gtrsim k_\perp  \gg \Lambda_{\rm QCD}$. Unlike the S-matrix that is Gaussian in $r_\perp^2Q_s^2(Y)$, this time it would not be trivial to evaluate the integral as, the Levin-Tuchin solution, having the following expression, 
\begin{eqnarray}
 S\left(r_{\bot}, Y\right) = \exp \left( - \tau \ln^2\left[r_\perp^2Q_s^2(Y)\right]\right)~, 
\label{LT}
\end{eqnarray}
\footnote{ with,
\begin{eqnarray}
\tau= \frac{1+2i\nu_0}{4\chi\left(0,\nu_0\right)} ~,
\end{eqnarray}
where the function $\chi$ is defined in terms of the digamma functions, 
\begin{eqnarray}
\chi\left(0,\nu\right) \equiv 2\psi(1) 
- \psi\left(\frac{1}{2}+i\nu\right) 
- \psi\left(\frac{1}{2}-i\nu\right)~.
\end{eqnarray}
Special point $\nu_0$ is so chosen that $\nu_0\equiv \nu_{sp}\left(r_\perp = 1/Q_s(Y),Y\right)\approx -0.1275 i$ which leads to $\tau \approx 0.2$. } is Gaussian in scaling variable, $\ln(r_\perp^2Q_s^2(Y))$ and valid when it is large $r_\perp^2Q_s^2(Y) \gtrsim 1$, leading to the large logarithm in the exponent, and hence, when expanded in series, can't be truncated at any finite order term in the series, to have a reasonable approximation.   \\

 %
 %A complete solution that span over full kinematic range of saturation dynamics is expected to be in accordance with both the 
 %McLerran-Venugopalan type initial condition and 
 %the Levin-Tuchin solution  in their appropriate limits.   \\  \\
%
  \begin{figure}
 %\includegraphics[width=0.4\linewidth]{fig1.pdf}\label{S_vs_xQ}
 %\caption{$S(x_{10},Y)$ as function of $x_{10}Q_s(Y)$} 
 %\includegraphics[width=0.6\linewidth]{Plot_G.eps}\label{N_vs_xQ}.
 % \includegraphics[width=0.6\linewidth]{Plot_K.eps}\label{N_vs_xQ}.
  \includegraphics[width=3.5in, angle=270]{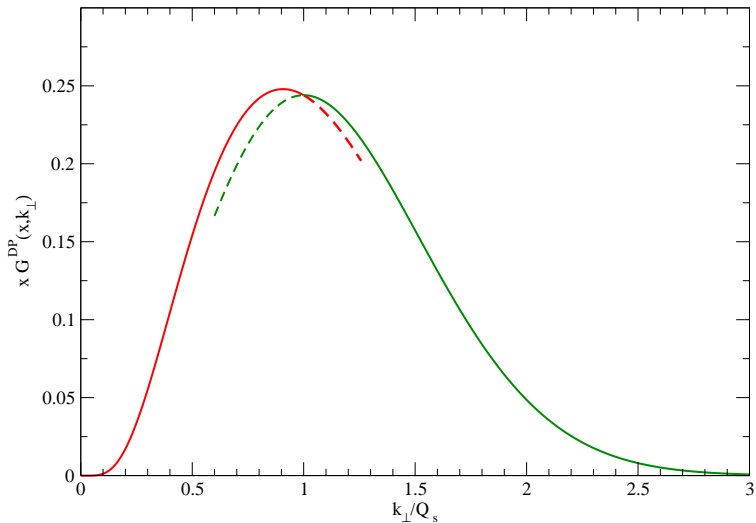}\label{N_vs_xQ}.
 \caption{The unintegrated dipole gluon distribution  $xG^{DP}(r_\perp,Y)$ plotted as function of $\xi=k_{\perp}/Q_s(Y)$ for nucleus of typical radius $\sim 7$ fm at $\alpha_s \sim 0.1$. Red line corresponds to Eq.(1) ($\propto \ln \xi ~\exp(-\tau \ln^2\xi)$) for the small momentum  range while green lines corresponds to Eq.(2) ($\propto \xi^2 \exp(-\xi^2)$) for relatively higher momentum. The tail of the green curve, has been tweaked by scaling down to match at $\xi =1$, about to follow power law fall as given in Eq.(\ref{xG-DP1001112}) and Eq.(\ref{xG-DP1001113}).}
 %The new solution, Eq.\eqref{AS}, compared with numerical solutions \cite{Albacete:2007yr} of running coupling improved leading order 
 %Balitsky-Kovchegove equation for both
 %Balitsky scheme \cite{Balitsky:2006wa} and Kovchegov-Wigert scheme \cite{Kovchegov:2006vj} with MV initial condition
 %(tweaked to reproduce $N=1/2$ at $\tau=1$).
 %McLerran-Venugopalan initial
 %condition Eq.\eqref{MV} and Levin-Tuchin solution Eq.\eqref{LT} also displayed for reference.} 
 %\label{SexyFigure1}
 %\end{figure}
 %
 \label{SexyFigure1}
 \end{figure}
%\section{unintegrated dipole gluon distribution functions at high energy}

We begin 
%with the caveat that expression of unintegrated dipole gluon distribution as defined Eq.\eqref{xG-DP100} may not be absolutely valid deep inside the saturation limit, however presuming that domain of validity for the expression of the distribution function in Eq.\eqref{xG-DP100} extends way to black disc limit, we write,  
%
by writing, 
\begin{eqnarray}
 \left. x G^{DP}\left(x,k_{\bot}\right)\right|_{Q_s \gtrsim k_\perp  \gg \Lambda_{\rm QCD}}^{}  &=& \frac{S_{\bot}N_c}{2\pi^2\alpha_{s}}~k_\perp^2  
 %%%%%%%%
 \int\frac{d^2r_{\bot}}{(2\pi)^2}e^{-ik_{\bot}.r_{\bot}} 
 %%%%%%%%
\exp \left( - \tau \ln^2\left[r_\perp^2Q_s^2(Y)\right]\right).
\label{xG-DPLT101} 
\end{eqnarray}
 Here we use the identity $\mathrm{ln}^2(y)=\mathrm{ln}^2(1/y)$, expand the exponential to 
express it in the form of a series where the $n$-th  term having $2n$-th derivatives of the dummy variable $\eta$ as, 
 \begin{eqnarray}
  \left. x G^{DP}\left(x,k_{\bot}\right)\right|_{Q_s \gtrsim k_\perp  \gg \Lambda_{\rm QCD}}
 %&=&\frac{S_{\bot}N_c}{2\pi^2\alpha_{s}}~k_\perp^2
 %\int\frac{d^2r_{\bot}}{(2\pi)^2}e^{-ik_{\bot}.r_{\bot}} \exp \left( - \tau \ln^2\left[r_\perp^2Q_s^2(Y)\right]\right)~,  \nn \\
 &=& \frac{S_{\bot}N_c}{2\pi^2\alpha_{s}}~k_\perp^2\sum_{n=0}^{\infty}
 \frac{(-\tau)^n}{n!} 
 \int\frac{d^2r_{\bot}}{(2\pi)^2}e^{-ik_{\bot}.r_{\bot}}  \ln^{2n}\left[\frac{1}{r_\perp^2Q_s^2(Y)}\right]~, \nn \\
 &=&\frac{S_{\bot}N_c}{2\pi^2\alpha_{s}}~k_\perp^2\sum_{n=0}^{\infty}
 \frac{(-\tau)^n}{n!} \lim_{\eta\rightarrow 0}\frac{\partial^{2n}}{\partial \eta^{2n}}\int \frac{d^2 r_{\bot}}{(2\pi)^2} 
~e^{-i k_{\bot}.r_{\bot}}\left[\frac{1}{r_{\bot}^{2}Q_{s}^{2}(Y)}\right]^{\eta} 
\label{xG-DP-02}
 \end{eqnarray}
This can further be simplified by changing the order summation and integration as long as the series converges and performing the integration for each term of the series, 
 \begin{eqnarray}
 \left. x G^{DP}\left(x,k_{\bot}\right)\right|_{Q_s \gtrsim k_\perp  \gg \Lambda_{\rm QCD}}^{} &=&\frac{S_{\bot}N_c}{2\pi^2\alpha_{s}}~k_\perp^2\sum_{n=0}^{\infty}
 \frac{(-\tau)^n}{n!} \lim_{\eta\rightarrow 0}\frac{\partial^{2n}}{\partial \eta^{2n}}\frac{1}{\pi}\frac{d}{dk_{\bot}^{2}}
\left(\frac{k_\perp^2}{4Q_{s}^{2}(Y)}\right)^\eta \frac{\Gamma(1-\eta)}{\Gamma(1+\eta)} ~.
\label{xG-DP-02009}
  \end{eqnarray}
 Besides $2n$-th $\eta$ differentiation  Eq.\eqref{xG-DP-02009} further contains additionally single first order $k_\perp^2$  derivative.  The details of the calculation  are in Appendix A.  One may now take the $k_\perp^2$ derivative out of all  $\eta$ derivatives and then apply the general Leibniz rule for $n$-th derivative of product of two functions to get, 
 \begin{eqnarray}
%{\cal D}  := 
%\frac{2\pi^2\alpha_{s}}{S_{\bot}N_c} 
 \left. x G^{DP}\left(x,k_{\bot}\right)\right|_{Q_s \gtrsim k_\perp  \gg \Lambda_{\rm QCD}}^{} %\frac{1}{k_\perp^2}
 &=&
   \frac{S_{\bot}N_c}{2\pi^3\alpha_{s}}~k_\perp^2 \sum_{n=0}^{\infty}
 \frac{(-\tau)^n}{n!} \lim_{\eta\rightarrow 0}\frac{d}{dk_{\bot}^{2}} \sum_{k=0}^{2n}
 \binom{2n}{k} \left[\frac{\partial^{2n-k}}{\partial \eta^{2n-k}} 
  \left(\frac{k_\perp^2}{4Q_{s}^{2}(Y)}\right)^\eta~\right]
  {\cal C}_{k}(\eta) ~, 
 \label{xG-DPseries}
 \end{eqnarray}
 where, for brevity, we define, 
 \begin{eqnarray}
 {\cal C}_{k}(\eta) = \frac{\partial^{k}}{\partial \eta^{k}
 } \frac{\Gamma(1-\eta)}{\Gamma(1+\eta)}~.
 \end{eqnarray}
 We first perform the $(2n-k)$-th order derivatives of $\eta$, 
\begin{eqnarray}
%{\cal D}  := 
%\frac{2\pi^2\alpha_{s}}{S_{\bot}N_c} 
 \left. x G^{DP}\left(x,k_{\bot}\right)\right|_{Q_s \gtrsim k_\perp  \gg \Lambda_{\rm QCD}}^{} 
%&=&%\frac{1}{k_\perp^2}&=&
 %  \frac{S_{\bot}N_c}{2\pi^2\alpha_{s}}~k_\perp^2 \sum_{n=0}^{\infty}
% \frac{(-\tau)^n}{n!} \lim_{\eta\rightarrow 0}\frac{d}{dk_{\bot}^{2}} \sum_{k=0}^{2n}
% \binom{2n}{k} \frac{\partial^{2n-k}}{\partial \eta^{2n-k}} 
%  \left(\frac{k_\perp^2}{4Q_{s}^{2}(Y)}\right)^\eta~
%  \frac{\partial^{k}}{\partial \eta^{k}
 %} \frac{\Gamma(1-\eta)}{\Gamma(1+\eta)}~, \nn \\
 %%%%%%%%%%%%%%%%%%
 &=&    \frac{S_{\bot}N_c}{2\pi^3\alpha_{s}}~k_\perp^2 \sum_{n=0}^{\infty}
 \frac{(-\tau)^n}{n!} \lim_{\eta\rightarrow 0}\frac{d}{dk_{\bot}^{2}} \sum_{k=0}^{2n}
 \binom{2n}{k} ~ {\cal C}_{k}(\eta) 
 \left(\frac{k_\perp^2}{4Q_{s}^{2}(Y)}\right)^\eta
 \ln^{2n-k}\left(\frac{k_\perp^2}{4Q_{s}^{2}(Y)}\right)~,\nn\\
 \end{eqnarray}
then the $k_\perp^2$ differentiation, 
\begin{eqnarray}
 \left. x G^{DP}\left(x,k_{\bot}\right)\right|_{Q_s \gtrsim k_\perp  \gg \Lambda_{\rm QCD}}^{} &=&%\frac{1}{k_\perp^2}&=&
   \frac{S_{\bot}N_c}{2\pi^3\alpha_{s}}~k_\perp^2\sum_{n=0}^{\infty}
 \frac{(-\tau)^n}{n!} \lim_{\eta\rightarrow 0}\sum_{k=0}^{2n}
 \binom{2n}{k} ~ {\cal C}_{k}(\eta) \left[
\left(\frac{\eta}{4Q_{s}^{2}(Y)}\right)\left(\frac{k_\perp^2}{4Q_{s}^{2}(Y)}\right)^{\eta-1}
\ln^{2n-k}\left(\frac{k_\perp^2}{4Q_{s}^{2}(Y)}\right) \right. \nn \\ 
 && \left. ~~~~~~~~~~~~~~~~~~~~~~~~~~~~~~~~~~~~~~~~~~~~~~~~~~+\left(\frac{k_\perp^2}{4Q_{s}^{2}(Y)}\right)^{\eta}\frac{d}{dk_{\bot}^{2}} 
  \ln^{2n-k}\left(\frac{k_\perp^2}{4Q_{s}^{2}(Y)}\right)\right]~, 
 \end{eqnarray}
and finally take the limit $\eta\rightarrow 0$, 
\begin{eqnarray}
%{\cal D}  := 
%\frac{2\pi^2\alpha_{s}}{S_{\bot}N_c} 
x G^{DP}(x,k_{\bot}) 
%&=&%\frac{1}{k_\perp^2}&=&
 %  \frac{S_{\bot}N_c}{2\pi^2\alpha_{s}}~k_\perp^2 \sum_{n=0}^{\infty}
% \frac{(-\tau)^n}{n!} \lim_{\eta\rightarrow 0}\frac{d}{dk_{\bot}^{2}} \sum_{k=0}^{2n}
% \binom{2n}{k} \frac{\partial^{2n-k}}{\partial \eta^{2n-k}} 
%  \left(\frac{k_\perp^2}{4Q_{s}^{2}(Y)}\right)^\eta~
%  \frac{\partial^{k}}{\partial \eta^{k}
 %} \frac{\Gamma(1-\eta)}{\Gamma(1+\eta)}~, \nn \\
 %%%%%%%%%%%%%%%%%%
% &=&    \frac{S_{\bot}N_c}{2\pi^2\alpha_{s}}~k_\perp^2 \sum_{n=0}^{\infty}
% \frac{(-\tau)^n}{n!} \lim_{\eta\rightarrow 0}\frac{d}{dk_{\bot}^{2}} \sum_{k=0}^{2n}
 %\binom{2n}{k} ~ {\cal C}_{k}(\eta) 
% \left(\frac{k_\perp^2}{4Q_{s}^{2}(Y)}\right)^\eta
% \ln^{2n-k}\left(\frac{k_\perp^2}{4Q_{s}^{2}(Y)}\right)~.\nn\\
 %%%%%%%%%%%%%%%%%%
% &=& \frac{S_{\bot}N_c}{2\pi^2\alpha_{s}}~k_\perp^2 \sum_{n=0}^{\infty}
% \frac{(-\tau)^n}{n!} \lim_{\eta\rightarrow 0}\sum_{k=0}^{2n}
% \binom{2n}{k} ~ {\cal C}_{k}(\eta) \left[
% \left(\frac{\eta}{4Q_{s}^{2}(Y)}\right)\left(\frac{k_\perp^2}{4Q_{s}^{2}(Y)}\right)^{\eta-1}
% \ln^{2n-k}\left(\frac{k_\perp^2}{4Q_{s}^{2}(Y)}\right) \right. \nn \\ 
% && \left. ~~~~~~~~~~~~~~~~~~~~~~~~~~~~~~~~~~~~~~~~~~~~~~~~~~+\left(\frac{k_\perp^2}{4Q_{s}^{2}(Y)}\right)^{\eta}\frac{d}{dk_{\bot}^{2}} 
%  \ln^{2n-k}\left(\frac{k_\perp^2}{4Q_{s}^{2}(Y)}\right)\right]~, \nn \\
 &=& \frac{S_{\bot}N_c}{2\pi^3\alpha_{s}}~k_\perp^2 \sum_{n=0}^{\infty}
 \frac{(-\tau)^n}{n!} \sum_{k=0}^{2n} \binom{2n}{k} ~ {\cal C}_{k}(0) \frac{d}{dk_{\bot}^{2}} 
  \ln^{2n-k}\left(\frac{k_\perp^2}{4Q_{s}^{2}(Y)}\right)~.
%   &=&\frac{S_{\bot}N_c}{2\pi^3\alpha_{s}}~k_\perp^2  \frac{d}{dk_{\bot}^{2}}  \sum_{n=0}^{\infty}
% \frac{(-\tau)^n}{n!} \sum_{k=0}^{2n} {\cal C}_{k}(0)\frac{1}{k!}
% \frac{\partial^{k}}{\partial \beta^{k}} \beta^{2n}~,
 \label{xG-DPseries1212}
 \end{eqnarray}
Defining, for convenience,  
 \begin{eqnarray}
 \lim_{\eta\rightarrow 0}  {\cal C}_{k}(\eta) = {\cal C}_{k}(0)
 ~~~\mathrm{and} ~~\beta=\ln\left(\frac{k_\perp^2}{4Q_{s}^{2}(Y)}\right)~, 
 \end{eqnarray}
 and noting that, 
 \begin{eqnarray}
 \beta^{2n-k}=\frac{(2n-k)!}{(2n)!}\frac{\partial^k}{\partial \beta^{k}} \beta^{2n}
 \end{eqnarray}
 Eq.\eqref{xG-DPseries1212} further simplified to, 
 \begin{eqnarray}
 \left. x G^{DP}\left(x,k_{\bot}\right)\right|_{Q_s \gtrsim k_\perp  \gg \Lambda_{\rm QCD}}^{}   &=&\frac{S_{\bot}N_c}{2\pi^3\alpha_{s}}~k_\perp^2  \frac{d}{dk_{\bot}^{2}}  \sum_{n=0}^{\infty}
\frac{(-\tau)^n}{n!} \sum_{k=0}^{2n} {\cal C}_{k}(0)\frac{1}{k!}
\frac{\partial^{k}}{\partial \beta^{k}} \beta^{2n}~,
 \end{eqnarray}
%Here we note that for a given $n$ the $k$-summation is non zero only upto $k=2n$ and terms having $k>2n$ vanish 
%as all the higher derivative terms are zero $i.e.$,
%\begin{eqnarray}
%\frac{\partial^{k}}{\partial \beta^{k}} \beta^{2n} = 0
%\end{eqnarray}
%for $k>2n$. 
%Hence, we may expand the upper limit of the summation way to infinity, 
%\begin{eqnarray}
%\sum_{k=0}^{2n} \Rightarrow \sum_{k=0}^{\infty}
%\end{eqnarray}
%We now interchange the order of b summations to get the following expression for the unintegrated gluon distribution, 
 %
% %
% where, 
%\begin{eqnarray}
% {\cal C}_{k}(0)=~\lim_{\eta \rightarrow 0}\frac{\partial^{k}}{\partial \eta^{k}
% } \frac{\Gamma(1-\eta)}{\Gamma(1+\eta)}~.
% \end{eqnarray}
% %
%The unintegrated dipole gluon distribution can then be written  in the form of double summation, 
 % \begin{eqnarray}
% x G^{DP}(x,k_{\bot})
 %&=&\frac{S_{\bot}N_c}{2\pi^3\alpha_{s}}~k_\perp^2  \frac{d}{dk_{\bot}^{2}}  \sum_{n=0}^{\infty}
 %\frac{(-\tau)^n}{n!} \sum_{k=0}^{2n} {\cal C}_{k}(0)\frac{1}{k!}
 %\frac{\partial^{k}}{\partial \beta^{k}} \beta^{2n}~,
 %\label{xG-DP-02}
%\end{eqnarray}
%where, for notational convenience, we defined, 
%\begin{eqnarray}
% \beta:=\ln\left(\frac{k_\perp^2}{4Q_{s}^{2}(Y)}\right)~.
% \end{eqnarray}
 We now interchange order of the summations, 
 \begin{eqnarray}
 \left. x G^{DP}\left(x,k_{\bot}\right)\right|_{Q_s \gtrsim k_\perp  \gg \Lambda_{\rm QCD}}^{} 
 &=&\frac{S_{\bot}N_c}{2\pi^3\alpha_{s}}~k_\perp^2  \frac{d}{dk_{\bot}^{2}}  
 \sum_{k=0}^{\infty}
 \sum_{n=\lceil \frac{k}{2} \rceil}^{\infty}
 \frac{(-\tau)^n}{n!}  {\cal C}_{k}(0)\frac{1}{k!}
 \frac{\partial^{k}}{\partial \beta^{k}} \beta^{2n}~,
 \label{xG-DP-03}
 \end{eqnarray}
 Here we note, 
 \begin{eqnarray}
 \sum_{n=0}^{\lceil \frac{k}{2} \rceil -1}
 \frac{(-\tau)^n}{n!}  {\cal C}_{k}(0)\frac{1}{k!}
 \frac{\partial^{k}}{\partial \beta^{k}} \beta^{2n}=0~,
 \label{xG-DP-04}
 \end{eqnarray}
 which helps to start the index of both the  summations from zero as, 
 \begin{eqnarray}
 \left. x G^{DP}\left(x,k_{\bot}\right)\right|_{Q_s \gtrsim k_\perp  \gg \Lambda_{\rm QCD}}^{} 
 &=&\frac{S_{\bot}N_c}{2\pi^3\alpha_{s}}~k_\perp^2  \frac{d}{dk_{\bot}^{2}}  
 \sum_{k=0}^{\infty}{\cal C}_{k}(0)\frac{1}{k!}
 \frac{\partial^{k}}{\partial \beta^{k}}
 \sum_{n=0}^{\infty}
 \frac{(-\tau)^n}{n!}   \beta^{2n}~,  
 \label{pompi}
 \end{eqnarray} 
 leading to exponentiation of the second series, 
 \begin{eqnarray}
 \left. x G^{DP}\left(x,k_{\bot}\right)\right|_{Q_s \gtrsim k_\perp  \gg \Lambda_{\rm QCD}}^{} 
  &=&\frac{S_{\bot}N_c}{2\pi^3\alpha_{s}}~k_\perp^2  \frac{d}{dk_{\bot}^{2}}  
 \sum_{k=0}^{\infty}{\cal C}_{k}(0)\frac{1}{k!}
 \frac{\partial^{k}}{\partial \beta^{k}}
  \mathrm{exp}(-\tau \beta^2)~, 
 \label{pompi}
 \end{eqnarray} 
 which can  further  be written as,  
  \begin{eqnarray}
  \left. x G^{DP}\left(x,k_{\bot}\right)\right|_{Q_s \gtrsim k_\perp  \gg \Lambda_{\rm QCD}}^{} 
 &=&\frac{S_{\bot}N_c}{2\pi^3\alpha_{s}}~k_\perp^2  \frac{d}{dk_{\bot}^{2}}  
 \mathrm{exp}(-\tau \beta^2)  \sum_{k=0}^{\infty} (-1)^{k}{\cal C}_{k}(0)\frac{\tau^{\frac{k}{2}}}{k!} H_{k}\left(\sqrt{\tau}\beta\right)~,
 \label{pompi}
 \end{eqnarray} 
%  where, 
% \begin{eqnarray}
% \beta=\ln\left(\frac{k_\perp^2}{4Q_{s}^{2}(Y)}\right)~.
% \end{eqnarray}
% In the last line we have used the formula, 
 where we use, 
 \begin{eqnarray}
 H_{k}(x)=(-1)^{k}e^{x^2}~\frac{d^k}{dx^k}~e^{-x^2}~. 
 \end{eqnarray} \\

The coefficients ${\cal C}_{k}(0)$ remains to evaluate.  In Appendix B we have  shown in detail that the coefficients ${\cal C}_{k}(0)$ can actually be expressed as the Bell's polynomials \cite{Bell:1927} \footnote{The complete exponential Bell polynomial $B_n(x_1,...,x_n)$ is defined by, 
 \begin{eqnarray}
\exp\left(\sum_{j=1}^{\infty}x_j \frac{t^j}{j!}\right)=\sum_{n=0}^{\infty}B_n(x_1,...,x_n)\frac{t^n}{n!}  \nn
\end{eqnarray}
The Bell's polynomial appear in the study of set partitions. The $n$-th Bell's polynomial is defined by $B_n(x_1,x_2,......x_n)=\sum_{k=1}^n B_{n,k}(x_1,x_2,......x_{n-k+1})$, where $B_{n,k}$ are the partial Bell's polynomials. Let's take an example: for $n=3, k=2$, $B_{3,2}(x_1,x_2)= 3 x_1 x_2$. $B_{3,2}$ tells us that we are considering partitioning a set of 3 elements into two boxes. $x_i$ indicates a box with $i$ elements, and the coefficient in front indicates the number of ways partitioning can be done.} 
 %*****************FOOTNOTE*****************
  of zeta functions for odd integers as, 
  \begin{eqnarray}
 {\cal C}_k(0)= B_k(x_1,...,x_k) := {\cal B}_k \nn   
%{\cal C}_k(0)={B}_k(2\gamma, 0, 2(2!)\zeta(3),0,2 (4!)\zeta(5),...) 
  \end{eqnarray}
 where, 
 \begin{eqnarray}
 x_1 &=&2\gamma~, \nn \\ 
x_2 &=& 0~, \nn \\
 x_3 &=&~2(2!)\zeta(3), \nn \\
x_4 &=& 0~, \nn \\
x_5 &=&~2(4!)\zeta(5), \nn \\
&...& \nn 
 \end{eqnarray}
 and $x_k=2(k-1)!\zeta(k)$ if $k$ is odd and $x_k=0$ if $k$ even with $\gamma$ being Euler-Mascheroni constant. 
%
%Henceforth, we replace ${\cal C}_{k}(0)$ by ${\cal B}_{k}(x_1,x_2,......x_k)\equiv {\cal B}_{k}$, where $x_1=2\gamma$ and $x_j=2~(j-1)!~\zeta(j)$ if $j$ is odd integer ($>$1) and zero if $j$ is even and continue with the calculation. Without going into further details of the Bell's polynomial, which have been deferred to Appendix B, \noindent Eq. \eqref{pompi} can now be written as, 
%
 \begin{eqnarray}
 \left. x G^{DP}\left(x,k_{\bot}\right)\right|_{Q_s \gtrsim k_\perp  \gg \Lambda_{\rm QCD}}^{} 
 &=&\frac{S_{\bot}N_c}{2\pi^3\alpha_{s}}~k_\perp^2  \frac{d}{dk_{\bot}^{2}}  
 \mathrm{exp}(-\tau \beta^2) \sum_{k=0}^{\infty} \frac{{\cal B}_{k}}{k!} ~\left({-\sqrt \tau}\right)^{k}H_{k}\left(\sqrt{\tau}\beta\right)~, \label{hottie}
  \end{eqnarray} 
 Interestingly, we also note that, 
\begin{eqnarray}
 B_{2n-k}\left[\ln \frac{k_\perp^2}{4Q_{s}^{2}(Y)},0,0,0,0,...\right]=\ln^{2n-k}\left(\frac{k_\perp^2}{4Q_{s}^{2}(Y)}\right)~, 
 \end{eqnarray}
 Then, taking the $k_{\perp}^2$ derivative out of the summation one may write Eq.\eqref{xG-DPseries1212} as, 
 \begin{eqnarray}
  &&  \left. x G^{DP}\left(x,k_{\bot}\right)\right|_{Q_s \gtrsim k_\perp  \gg \Lambda_{\rm QCD}}^{}  \nn \\
  &=& \frac{S_{\bot}N_c}{2\pi^2\alpha_{s}}~k_\perp^2  \sum_{n=0}^{\infty}
 \frac{(-\tau)^n}{n!}\frac{d}{dk_{\bot}^{2}}  \sum_{k=0}^{2n} \binom{2n}{k} ~ B_k\left[2\gamma,0,2(2!)\zeta(3),0,2(4!)\zeta(5),0,...\right] ~
 B_{2n-k}\left[\ln \frac{k_\perp^2}{4Q_{s}^{2}(Y)},0,0,0,0,...\right]~, \nn \\
  &=& \frac{S_{\bot}N_c}{2\pi^2\alpha_{s}}~k_\perp^2 \frac{d}{dk_{\bot}^{2}}  \sum_{n=0}^{\infty}
 \frac{(-\tau)^n}{n!}   B_{2n}\left[\ln \frac{k_\perp^2}{4Q_{s}^{2}(Y)}+2\gamma , 0,1!~2 \zeta(3),0,4!~2\zeta(5),..,(2n-2)!~2\zeta(2n-1),0\right] ~.  \label{almostfinal}
 \end{eqnarray}

\subsection{Leading log resummation}

For $k_\perp^2 \ll Q_s^2$ one may approximate the Bell's polynomials as, 
 \begin{eqnarray}
  B_{2n}\left[\ln \frac{k_\perp^2}{4Q_{s}^{2}(Y)}+2\gamma , 0,1!~2 \zeta(3),0,4!~2\zeta(5),..,(2n-2)!~2\zeta(2n-1),0\right]  \approx
  \left(\ln \frac{k_\perp^2}{4Q_{s}^{2}(Y)}+2\gamma\right)^{2n}
 \end{eqnarray}
%
% Above Bell's polynomial is essentially a polylogarithmic function in $k_\perp^2/Q_s^2$ in the form of,
% \begin{eqnarray}
% B_{2n} \equiv a_k \ln ^{2n}\frac{k_\perp^2}{4Q_{s}^{2}(Y)}+... + a_1 \ln \frac{k_\perp^2}{4Q_{s}^{2}(Y)}+a_0~.
% \end{eqnarray}
 % 
 and the unintegrated gluon distribution can be approximated (to the leading log) as:
   \begin{eqnarray}
 \left. x G^{DP}\left(x,k_{\bot}\right)\right|_{Q_s \gg k_\perp  \gg \Lambda_{\rm QCD}}^{} 
 &= &  
  \frac{S_{\bot}N_c \tau}{2\pi^3\alpha_{s}} k_\perp^2  \frac{d}{d k_\perp^2}\exp \left[-\tau \left(\ln\left(\frac{k_\perp^2}{4Q_s^2}\right) + 2\gamma\right)^2\right] , \\
&=& -\frac{S_{\bot}N_c \tau}{\pi^3\alpha_{s}}~\left[\ln\left(\frac{k_\perp^2}{4Q_s^2}\right)+2\gamma \right] \exp \left[-\tau \left(\ln\left(\frac{k_\perp^2}{4Q_s^2}\right) +2\gamma\right)^2\right]~,   \\ 
&\approx& -\frac{S_{\bot}N_c \tau}{\pi^3\alpha_{s}}~\ln\left(\frac{k_\perp^2}{4Q_s^2}\right) \exp \left[-\tau \ln^2\left(\frac{k_\perp^2}{4Q_s^2}\right) \right]~. 
 \end{eqnarray} 
%Above expression posses striking similarity with the Sudakov form factor, when %summing double logarithms, upto one prefactor in the exponent of the soft factor.  
% All polylogarithmic functions of $k_\perp^2/Q_s^2$ are $o(k_\perp^{2\epsilon}/ Q_s^{2\epsilon})$ for every exponent $\epsilon >0$, with symbol `o' being small o notation. The difference is in whether the two functions may be asymptotically same or   not. While the big-O meaning ``grows no faster than' ' ($i.e.$ grows at the same rate or slower) and little-o meaning ``grows strictly slower than''. 

Inside the saturation region, the dipole gluon distribution is expected to go to zero as $k_\perp^2 \rightarrow 0$. This is indeed the case here, however, 
interestingly, we observe a Sudakov double logarithm type factor in the argument of the exponential function and also, at small transverse momentum, $x G^{DP}(x,k_{\bot})$ not proportional to $k_\perp^2$, as previously anticipated \cite{Kharzeev:2003wz,Buffing:2013eka}, rather it is proportional to $\ln\left(k_\perp^2/4Q_s^2\right)$ times the double log soft factor. We note here until Eq.(\ref{almostfinal}) we are carrying the $k_\perp^2$ (and also one $d/dk_\perp^2$, origin of which can be traced back to Appendix A).
 The $k_\perp^2$ is been killed when we perform the summation
 in Eq.(\ref{almostfinal}), in leading log approximation, and then operate the derivative, to reach our final result . After operating the $d/dk_\perp^2$ on the exponential it generates one $1/k_\perp^2$ that kills the$k_\perp^2$ in the numerator. This indicates that  evolution kills the $k_\perp^2$ behaviour  from Eq.(\ref{xG-DP1001111}) to Eq.(\ref{main1234}) in the small momemtum limit and modifies it to be a {\it log} $\bigotimes$ {\it log normal} distribution as given in Eq.(1) which is the main finding of this work. 

%The variation of the scaled leading log approximated gluon distribution with respect to the scaled gluon transverse momentum square $\tilde{k}_{\bot}^2=
%k_{\bot}^2/Q_s^2$ has been plotted in Fig.~\ref{1}. 
%\

%begin{figure}[ht]
%\begin{center}%minipage}[h]{0.3\textwidth}
%\includegraphics[scale=0.74]{Distunint.pdf}
%\end{minipage} \hspace{0.2\textwidth}
%\begin{minipage}[h]{0.3\textwidth}
%\includegraphics[scale=0.34]{diffperptemp.pdf}
%\end{center}%minipage}
%\caption{Variation of the scaled unintegrated gluon distribution in the leading log approximation for $\tau=0.2$.}
%\label{1}
%\end{figure}

\section{Conclusion and Outlook}

 Classical asymptotic analysis covers the study of the limiting behaviour of functions when some special (singular or non singular) points are approached.  
\noindent  Asymptotic expansions usually give increasingly better approximations as the special points are approached, yet not necessarily they always converge to the  actual function.  
In this paper we derive results for  unintegrated color dipole gluon distribution function at small transverse momentum.
By Fourier transforming the $S$-matrix for large dipoles
%as given by the Levin-Tuchin solution of the Balitsky-Kovchegov equation, to transverse momentum space, 
we derive the results in the form of a  series of Bells polynomials that comes in the study of combinatorics. 
Interestingly, when resumming  the series in leading log accuracy, the results showing up striking similarity with the Sudakov form factor with role play of coupling is being done by a constant that stems from the saddle point condition along the saturation line.   
\noindent  In QCD non-perturbative effects, within the framework of perturbative analysis, stems from the asymptotic nature of the perturbation series having coupling as the expansion parameter.  \noindent In this context it would be interesting to see how non-perturbative  renormalon contribution shows up when including running coupling effects in this study. 
%
%\noindent Here the small-$x$ evolution effects are included by computing the unintegrated dipole gluon distribution function exclusively from the small-$x$ asymptotic solution, of dipole scattering amplitudes, that obeys the  Balitsky Kovchegov equation therefore only resumes small-$x$ logarithms and not the Sudakov type large logarithms \cite{Mueller:2012uf}. 
%
 It would be also interesting to see how this conncets with CSS evolution of TMDs \cite{Collins:1981uk,Collins:1984kg}.  
 As $S$-matrix directly connects with the probability of transverse deflection of the interacting parton this result  could therefore possibly be useful to study medium sensitive observables such as jet quenching parameter \cite{Abir:2015qva}. We are also working on to see how the result modify when taking, the recently derived analytical, solution that valid both deep in and way out of the saturation region \cite{Abir:2017mks} and why leading log approximation of the solution posses similarity with the Sudakov soft factor \cite{Liu:2017vkm,Penin:2014msa}.

 \begin{acknowledgments}
 %{\it Acknowledgments~:~} 
 %We are indebted to Yuri Kovchegov for many valuable suggestions and comments since inception of this work. 
 %We also thank Rafi Alam, Trambak Bhattecharya, Haider H. Jafri and Manjari Sharma for valuable discussions and help.
We indebted to Yuri Kovchegov for important suggestions on this work. We also thank Bo-Wen Xiao, Kirill Tuchin, Cyrille Marquet for valuable comments on the draft. 
This work was supported in part by the University Grants Commission under UGC-BRS Research Start-Up-Grant grant number F.$30$-$310/2016$(BSR). 
T.B. acknowledges the University Research Committee, University of Cape Town, South Africa for support. 
 \end{acknowledgments}

%**********************************************************************%**********************************************************************
%**********************************************************************%**********************************************************************

\section*{Appendix A}

For circularly symmetric function two dimensional Fourier transform is essentially 
the Hankel transform of order zero, 
 \begin{eqnarray}
 I &:=&\int \frac{d^2 r_{\bot}}{(2\pi)^2} ~e^{-i k_{\bot}.r_{\bot}}\left[\frac{1}{r_{\bot}^{2}Q_{s}^{2}(Y)}\right]^{\eta} 
 = \frac{1}{2\pi}Q_{s}^{-2\eta}(Y)\int_{0}^{\infty} d r_{\bot}r_{\bot}J_{0}(r_{\bot}k_{\bot})r_{\bot}^{-2\eta} ~.
 \label{Hankel}
 \end{eqnarray}
 The integral in Eq.\eqref{Hankel} can be evaluated using following integral identities of Bessels functions,  
 \begin{eqnarray}
 \int_{0}^{\infty} dk k^{\lambda-1} J_0\left(kx\right)
 =2^{\lambda-1}x^{-\lambda}\frac{\Gamma\left(\lambda/2\right)}{\Gamma\left(1-\frac{\lambda}{2}\right)} ~.  \label{hot}
 \end{eqnarray} 
 Therefore, 
 \begin{eqnarray}
 I&=&\frac{\eta}{\pi}~\left(\frac{1}{4Q_{s}^{2}(Y)}\right)^\eta \frac{\Gamma(1-\eta)}{\Gamma(1+\eta)}
 ~ \left(k_{\bot}^2\right)^{\eta-1} ~, 
 \end{eqnarray}
 which may then further be simplified to, 
\begin{eqnarray}
 I&=& \frac{1}{\pi}~\frac{d}{d k_\perp^2}~\left(\frac{k_\perp^2}{4Q_{s}^{2}(Y)}\right)^\eta \frac{\Gamma(1-\eta)}{\Gamma(1+\eta)}~.
 \label{Cinderella}
 \end{eqnarray}
 Unintegrated dipole gluon distribution can then be written as, 
  \begin{eqnarray}
\left. x G^{DP}(Y\equiv \ln \frac{1}{x},~k_{\bot})\right|_{} 
 &=&\frac{S_{\bot}N_c}{2\pi^2\alpha_{s}}~k_\perp^2\sum_{n=0}^{\infty}
 \frac{(-\tau)^n}{n!} \lim_{\eta\rightarrow 0}\frac{\partial^{2n}}{\partial \eta^{2n}}\frac{1}{\pi}\frac{d}{dk_{\bot}^{2}}
\left(\frac{k_\perp^2}{4Q_{s}^{2}(Y)}\right)^\eta \frac{\Gamma(1-\eta)}{\Gamma(1+\eta)} ~.
\label{xG-DP-03}
 \end{eqnarray}

 \section*{Appendix B: Evaluation of ${\cal C}_{k}$} 
 When $|z|<1$  the function $\Gamma (1+z)$ can be written as
 \begin{eqnarray}
 \Gamma (1+z) = \exp \left(-\gamma z + \sum_{j=2}^{\infty}(-1)^j\zeta(j)
 \frac{z^j}{j}\right)
 \end{eqnarray}
 Therefore, 
 \begin{eqnarray}
 \frac{\Gamma (1-\eta)}{\Gamma(1+\eta)} = \exp\left(2\gamma \eta+2 \sum_{j=1}^{\infty}\zeta(2j+1)
 \frac{\eta^{2j+1}}{2j+1}\right)
 \end{eqnarray}
 Hence the  coefficients ${\cal C}_{k}$ could be written as Bell's 
 polynomials for Riemann zeta function at odd integers as,  
 \begin{eqnarray}
{\cal C}_{k}= B(x_1,x_2,x_3,...,x_j,...,x_k)
\end{eqnarray}
where $x_1=2\gamma$ and $x_j=2~(j-1)!~\zeta(j)$ if $j$ is odd integer and zero if even. Initial a few 
coefficients are as follows, 
\begin{eqnarray}
{\cal C}_{0}(0) &=& 1 ~, \nn \\
{\cal C}_{1}(0) &=& B\left[{2\gamma}\right] = 2\gamma~, \nn \\
{\cal C}_{2}(0) &=& B\left[{2\gamma, 0}\right]=(2\gamma)^2~, \nn \\ 
{\cal C}_{3}(0) &=& B\left[{2\gamma, 0, 2(2!)\zeta(3)}\right]=(2\gamma)^3-2\psi(2,1)~, \nn \\
{\cal C}_{4}(0) &=& B\left[{2\gamma, 0, 2 (2!)\zeta(3),0}\right]=(2\gamma)^4-16\gamma\psi(2,1)~, \nn \\
{\cal C}_{5}(0) &=& B\left[{2\gamma, 0, 2(2!)\zeta(3),0,2 (4!)\zeta(5)}\right]=(2\gamma)^5-80\gamma^2\psi(2,1)-2\psi(4,1)~,
\end{eqnarray}
where $\psi$ is polygamma function.

\end{document}